# Copper Titanates: A Family of Magneto-Quantum Paraelectrics

Jitender Kumar and A.M. Awasthi*

Thermodynamics Laboratory, UGC-DAE Consortium for Scientific Research,

University Campus, Khandwa Road, Indore- 452 001, India

*amawasthi@csr.res.in

A magneto-quantum-paraelectric family is emergent from the shared low-temperature characteristics of representative $SrCu_3Ti_4O_{12}$ (SCTO), $CaCu_3Ti_4O_{12}$ (CCTO), and $Ca_{0.9}Li_{0.1}Cu_3Ti_4O_{12}$ (CLCTO) antiferro-tilted/$A_{1/4}A'_{3/4}BO_3$-structures. Above their magnetic ordering temperatures $T_N$, dielectric permittivity of SCTO and CLCTO follow typical quantum paraelectric Barrett-form unambiguously, whereas in CCTO, this behaviour is masked by the huge $\varepsilon'$-step, born of thermally-activated site-antisite (Ca/Cu) disorder. The *hidden* quantum paraelectricity in CCTO is revealed with Li-doping at Ca-site, by considerable temperature-scale-upshift of the colossal dielectric constant (CDC) anomaly.

The Perovskite $ABO_3$ structure is versatile and robust. It can be cubic, tetrahedral, or orthorhombic at standard temperature and pressure. In materials science Perovskites have great importance because of a huge variation in their properties such as in CMR [1], superconductivity [2], and most importantly in their dielectric properties [3]. $CaTiO_3$ is the father of all the Perovskite structures [4]. The distortions from perfect cubic/orthorhombic structure endow them with many important and fascinating phenomena viz., ferroelectricity, quantum paraelectricity, huge piezoelectricity etc. The main three types of distortions from the perfect cubic are ferroelectric (FE-- due to the inversion-symmetry breaking and net relative displacement between anions and cations), antiferro-distortive (AFD-- due to the rotation of the $Ti-O_6$ octahedra), and antiferro-tilted (AFT-- due to the tilting of the $Ti-O_6$ octahedra). The most famous example of FE is $BaTiO_3$, of AFD is $SrTiO_3$, and of AFT is the $CaTiO_3$ prototype. Here, $BaTiO_3$ undergoes more than one ferroelectric transition whereas the last two exhibit none, but rather feature quantum paraelectricity (QP). In QPs, the long-range FE is suppressed by the quantum/zero-point fluctuations, and the materials do not show any electrical ordering down to absolute zero. $SrTiO_3$ is a prototype quantum paraelectric [5]. Nowadays it is a general opinion that the tilting/rotation of the octahedral units are responsible for the suppression of the FE ground state. Here we report recognizing the family $ACu_3Ti_4O_{12}$ (A=Sr, Ca, Ba, Cd, $Na_{1/2}Bi_{1/2}$ etc. having $A_{1/4}A'_{3/4}BO_3$ double-Perovskite cubic structure with space group $Im3$) as also exhibiting the quantum paraelectric character, because of the large tilting of their $Ti-O_6$ octahedra. In the $ACu_3Ti_4O_{12}$ family the A'-site is occupied by the magnetic atom Cu; its presence responsible for the emergence of the quantum paraelectric glass (QPG) state [6] near the magnetic ordering temperature. Also, the signatures of incipient ferroelectricity are observed in [7].

The $ACu_3Ti_4O_{12}$ family was shot to fame by $CaCu_3Ti_4O_{12}$, which has colossal dielectric constant [8] over broad temperature and frequency ranges, due to the site-antisite disorder of Ca and Cu [9]. Its permittivity shows large drop on lowering the temperature below ~100K, without any ferroelectric/structural or relaxor-like transitions down 35K [10]. Yet, CCTO does not feature quantum paraelectricity; the huge $\varepsilon'$-step masking the same. The hidden QP nature of CCTO may be recoverable if the energy-scale of the activated site-antisite disorder can be temperature-upshifted. Here we report two compositional-derivatives having different levels of this disorder, and demonstrate quantum paraelectricity as the inherent characteristic of this family. $A_{1/4}A'_{3/4}BO_3$ family has a unique structure in which A-site is occupied by Ca, Sr, Ba etc. and A' is a magnetic atom (Cu), which undergoes antiferromagnetic ordering around 25K [11]. The tilted $Ti-O_6$ octahedra here are the same as in $CaTiO_3$ (CTO). However, the tilting in the CCTO family is more in comparison to that in CTO, because of the formation of planar $Cu-O_4$ rigid units. To demonstrate the QP nature of the $A_{1/4}A'_{3/4}BO_3$, we have synthesized three specimens viz., 1) CCTO, which has large degree of disorder and has a huge dielectric constant. 2) Li-doped CCTO having somewhat modes scale of disorder, and 3) SCTO, which does not have any site-antisite disordering. The dielectric spectra of these compositions reveal the common QP-nature of this family.

All the samples were prepared by the conventional solid state reaction method. The preparation of CCTO and SCTO are reported elsewhere [6,12]. $Ca_{0.9}Li_{0.1}Cu_3Ti_4O_{12}$ was made with high-purity (4N) $Li_2O_3$, $CaCO_3$, CuO, and $TiO_2$. The ingredients were mixed in stoichiometric amount and thoroughly hand ground. The finely-ground charge was first calcined at 950°C for 12 hrs, again ground for another 10 hrs. After sintering at 1000°C for 12 hrs pellets were made and given final heat treatment at 1000°C for 24 hrs. The pelletized specimens (10mm diameter and 1-3mm thick) were sintered again at 1100°C and silver-coated for good electrical contacts for the dielectric measurements. XRD measurements were carried out using Bruker D8 Advance X-ray diffractometer. The X-rays were produced using a sealed tube and the wavelength used was 1.54Å. Dielectric measurements over 5K to room temperature were performed using NOVO-CONTROL (Alpha-A) High Performance Frequency Analyzer across 0.5Hz to 10MHz, using 1V ac signal for excitation.

In figure (1) upper panel the dielectric permittivity and the dielectric loss of three different samples are shown. Associated with the $\varepsilon'$-step feature, relaxation peaks in the dielectric loss tangent are also observed (figure 1, lower panel). As per the mean-field theoretical calculations [13], permittivity of CCTO should be less than ~100 at room temperature, but experimentally it is observed order of magnitudes higher, and is almost constant over a wide temperature window. The high dielectric constant is well studied and is understood as due to the nano-scale site-antisite/charge disorder of Ca- and Cu-cations, which makes a platform for the inter-barrier layer capacitance (IBLC) mechanism [14]. The drop on lowering the temperature is due to the anti-parallel correlations of the relaxing entities [9]. We have down-tuned the disorder in CCTO by Li-doping at the Ca-site, which favourably increases the activation energy vs. that in CCTO, responsible for the permittivity-step feature. In the Li-doped CCTO, $\varepsilon'$-step shown for the 100Hz and 1kHz data is upshifted to near the room temperature, the same being below ~100K for the pure CCTO masks its intrinsic low-temperature QP nature. The dielectric behaviour of SCTO makes the picture still clearer, which does not have any step-feature up to the room temperature. The value of the dielectric constant of SCTO is below ~100 at room temperature, which is quite matched with the theoretical value (~70) for this structure, making sure that this composition does not have any site-antisite disorder. The quantum paraelectric nature of SCTO is already demonstrated in an earlier published report [6]. The intrinsic behaviour of this sample clearly appears as a quantum paraelectric, because its permittivity has the negative temperature coefficient and follows the Barrett temperature dependence typical of QP's [15].

In figure (2) we show the X-Ray diffraction of the Li-doped CCTO with Rietveld refinement of the LCTO to make sure the sample is single phase by using the full-prof software by taking crystal symmetry $Im3$. No extra reflections ensure the good quality of our samples. The chi-square and goodness of the fit are 2.56 and 1.6 respectively. Comparison of the lattice parameters, bond lengths, and bond angles of the three samples (LCTO, CCTO, and SCTO) is as tabulated below.

| Sample | Lattice Parameter | Ti-O-Ti Bond-angle | Cu-O Bond-length | Ca-O Bond-length | Ti-O Bond-length |
|---|---|---|---|---|---|
| $CaCu_3Ti_4O_{12}$ | $a=b=c=7.394$Å | 143.40° | 2.013 Å | 2.581 Å | 1.946 Å |
| $Ca_{0.9}Li_{0.1}Cu_3Ti_4O_{12}$ | $a=b=c=7.362$Å | 141.58° | 1.9743Å | 2.58215 Å | 1.950 Å |
| $SrCu_3Ti_4O_{12}$ | $a=b=c=7.405$Å | 141.83° | 1.978 Å | 2.666 Å | 1.958 Å |

Magneto quantum paraelectricity in this family is rooted in their common crystal structure, which is based on the Perovskite building-block. Here, the unit cell comprises of eight (2 x 2 x 2) $ABO_3$ basic units, wherein the A-sites have the configuration $A_{1/4}A'_{3/4}$. In each $ABO_3$ sub-cell [9] two (body-diagonal) of the eight A-sites are occupied by the non-magnetic Ca/Sr-atoms and the remaining six by the magnetic Cu-atoms; their four different relative configurations possible (corresponding to the total of four body-diagonals) require eight $ABO_3$ sub-cells for the (*x-y-z* symmetrical) periodicity, to constitute a unit cell. The A-O framework in $ABO_3$ Perovskites evokes an interstitial space, which is larger than the actual size of the central $Ti^{4+}$ ion. As a result in the $ATiO_3$ (basic-Perovskites), a series of phase transformations takes place to reduce the Ti-cavity size. Certainly, the radii of the ions involved impact the propensity for the ferroelectric phase-formation; thus while both $PbTiO_3$ and $BaTiO_3$ have ferroelectric phases, $CaTiO_3$ and $SrTiO_3$ do not. In the present $A_{1/4}A'_{3/4}BO_3$ copper-titanates, the size of the A-O framework (2.581Å) is the same as that in the well-known quantum paraelectrics $SrTiO_3$ and $CaTiO_3$, thus providing the structural basis for the quantum paraelectricity in this family. The presence of magnetic (Cu) atom in the $A_{1/4}A'_{3/4}BO_3$ copper-titanates make them very special; till now only one other material ($EuTiO_3$) has been reported having both quantum paraelectricity and magnetic ordering in its undoped form [16].

For the confirmation of quantum paraelectricity in the copper-titanates family, we have fitted the dielectric permittivity by the Barrett form [15]; the formulation for the QP-behaviour given below.

$$\varepsilon'(T) = A + \frac{C}{\left[\left(\frac{T_1}{2}\right)\coth\left(\frac{T_1}{2T}\right) - T_C\right]}$$

Here $T_C$ is the Curie temperature and $T_1$ is the temperature below which the quantum fluctuations dominate, $A$ is the high-temperature permittivity-baseline, and $C$ is constant. The Barrett formula quite fits the dielectric permittivity (100kHz) of SCTO from room temperature down to the magnetic ordering at $T_N$, figure (3). The relevant fitting parameters for SCTO are $A = 44.06$, $T_1 = 158.41$K, and $T_C = -64.56$K. For LCTO, Barrett formula is not fittable up to the room temperature as above ~120K emergence of other processes affects the permittivity-behaviour. Barrett-parameters obtained for 1MHz permittivity data of CLCTO (figure (4)) are $A = 40.67$, $T_1 = 160.01$K, and $T_C = -81.51$K. Negative values of $T_C$ indicate the presence of antiferroelectric dipole-interactions in both the samples. Deviation of the data from the exact Barrett fit at lower temperatures (observed permittivity-drop as against its QP-expected level-off) is due to the interruption caused by the Cu-spins' AFM ordering at $T_N = 23$K. Strong spin-phonon coupling owing to the special structure of this family provides large magneto-electric (ME) effect, that tends to organize the material electrically, concurrent with the AFM order. However, the quantum fluctuations (QF) subvert this organization from achieving a long-range character down to 0K. As a compromise, medium range electrically ordered state is settled for, by the competing QF and ME influences.

At high temperatures, permittivity of any (classical as well as quantum) paraelectric must follow the Curie-Weiss (C-W) behaviour [~ $(T-T_C)^{-1}$]. The Barrett form is expected to merge into the C-W dependence when quantum fluctuations are overtaken by the thermal fluctuations at high temperatures. This can be anticipated from the asymptotic behaviour $\left[\left(\frac{T_1}{2}\right)\coth\left(\frac{T_1}{2T}\right) \xrightarrow{T \gg T_1} T\right]$ of the $T$-dependent part of the Barrett formula. However, to *exactly* visualize this simplification (i.e., the vanishing electrical susceptibility as $T \to \infty$), one *has to but remove* the high-temperature (constant) offset ($A$) from the Barrett-fit and the data, and then invert *only* their temperature-dependent part. Thus, when the offset-subtracted $(\varepsilon'-A)^{-1}$ is plotted versus the temperature, the typical $T$-linear C-W high-temperature behaviour *indeed* becomes obvious, as in figure (3). For SCTO, we notice this high-temperature Curie-Weiss linearity down to ~172K (within 8% of the Barrett-fitted $T_1 \approx 158$K), confirming the interpretation (as also previously reported, [6]) of $T_1$ as indicating the quantum-classical crossover temperature. Moreover, we also find $T_{C-W} = -79.89$K as somewhat close to the Barrett $T_C = -64.56$K. In the case of CLCTO, Barrett form is seen to apply onto the data only up to 120K, above which other processes deviate the permittivity from the fitted QP-dependence. Yet, here too down to ~167K (within 5% of the Barrett $T_1 \approx 160$K) the inverted (offset-subtracted) QP-Barrett fit $(\varepsilon'-A)^{-1}$ identifies at high temperatures with the $T$-linear C-W behaviour ($T_{C-W} = -97.11$K, within 16% of the Barrett $T_C = -81.51$K, figure (4)).

In figure (5) we show the low-temperature specific heat $C_p(T)$ and magnetization $M(T)$ of CLCTO. The clear magnetic transition is observed at $T_N = 23$K in both the data, same as for the other family members viz. SCTO [6] and CCTO [12]. The G-type AFM ordering of the Cu-spins here is mediated through the $Ti^{+4}$-cations rather than the oxygens (usual with other Perovskites), which makes for a more direct and appreciable magneto-dielectricity. Strong spin-phonon coupling is assured by the atypical structure of their unit cell. The oxygens of the $Ti-O_6$ octahedra are also directly bond to the Cu-atoms in $Cu-O_4$ square-planer arrangements; the latter forming a propeller-like configuration with the octahedron, as shown in figure 6. The signatures of phonon-softening near the AFM-ordering are reported in SCTO [6] and CCTO [11]. As the spin-phonon coupling is already established in this family, so is an expected change in their dielectric properties at the $T_N$. Figure (7) shows clear ($\omega$-$T$ dispersive) drops in CLCTO-permittivity near $T_N$. The kinetic character of this 'transition' can be understood, since this family has a unique combination of quantum fluctuations and spin-phonon driven magneto-dielectricity. As the competition product of these opposing influences, an electrically medium-range organized glassy state is realized in the system [6], responsible for their frequency-temperature dispersed permittivity near $T_N$. The $\varepsilon'$-peak frequencies here are seen to follow the Vogel-Fulcher-Tamman (VFT) dependence [17, 18] (inset, figure(7)), with freezing temperature $T_{VF}$ ~15K and the activation energy $E_a$ ~ 6.7 meV. This dispersion kinetics already reported in SCTO led to establishing a quantum paraelectric glass (QPG) state in the material [6]. The common characteristics therefore make compelling evidence for these copper-titanates as belonging to a family of magneto-quantum paraelectric glasses (MQPG).

In conclusion, we have demonstrated the shared quantum paraelectric (QP) manifestation of the family of double Perovskites $A_{1/4}A'_{3/4}BO_3$. The hidden QP character in $CaCu_3Ti_4O_{12}$ is unmasked by the Li-doping, which alters the nano-scale disorder and upshifts the giant $\varepsilon'$-step to near room temperature, revealing its underlying quantum paraelectric behaviour at the mid-low temperatures. Characteristic Barrett-permittivity behaviour down to $T_N$ and the magnetic-ordering driven QPG-state formation in both $SrCu_3Ti_4O_{12}$ and $Ca_{0.9}Li_{0.1}Cu_3Ti_4O_{12}$ confirm quantum paraelectricity as the common defining family-bond of the copper-titanates.

**Acknowledgements**
The authors thank Rajeev Rawat, R.J. Choudhary, and Mukul Gupta for heat capacity, magnetization, and XRD measurements respectively.

**Figure Captions**

Figure(1): Real permittivity (upper) and loss tangent (lower) vs. temperature of the three different members of the Copper-Titanates family. A remarkably large shift of the plateau towards higher temperatures on Li-doping of the CCTO specimen is of great practical interest.

Figure(2). Rietveld refined room temperature XRD pattern of $Ca_{0.9}Li_{0.1}Cu_3Ti_4O_{12}$.

Figure(3). Barrett fit on $SrCu_3Ti_4O_{12}$. Right $y$-axis: indistinguishability of Barrett fit $(\varepsilon'-A)^{-1}$ from the Curie-Weiss $(T-T_{C-W})$ linearity at high temperatures.

Figure(4). Barrett Fit on $Ca_{0.9}Li_{0.1}Cu_3Ti_4O_{12}$. Right $y$-axis: indistinguishability of Barrett fit $(\varepsilon'-A)^{-1}$ from the Curie-Weiss $(T-T_{C-W})$ linearity at high temperatures.

Figure(5). Heat-capacity and magnetization (inset) data of $Ca_{0.9}Li_{0.1}Cu_3Ti_4O_{12}$. The antiferromagnetic phase transition occurs at $T_N$=23K.

Figure(6). Each oxygen of Ti-$O_6$ octahedron also directly connects to Cu-atoms and forms Cu-$O_4$ square-planer substructures, which are arranged in propeller-like configuration with the central Ti-$O_6$ tetrahedron; providing a platform for strong spin-phonon coupling.

Figure (7). Glassy dispersion of the dielectric constant in the vicinity of $T_N$. Inset: Arrhenius-plot of probing frequency vs. inverse of $\varepsilon'$-peak temperature ($1/T_p$) fits the Vogel-Fulcher glassy slowdown with cooling.

Figure (1)

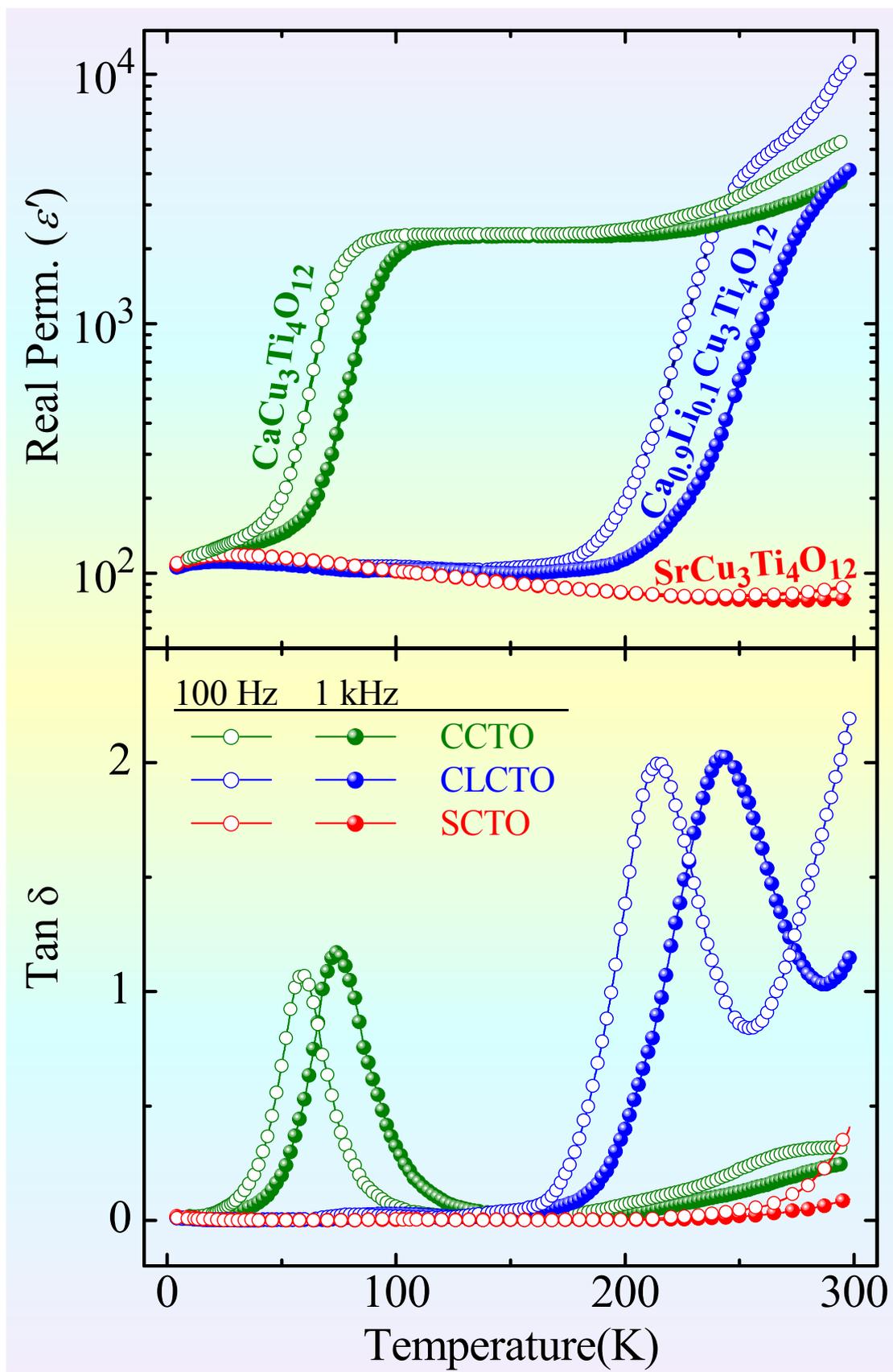

Figure(2)

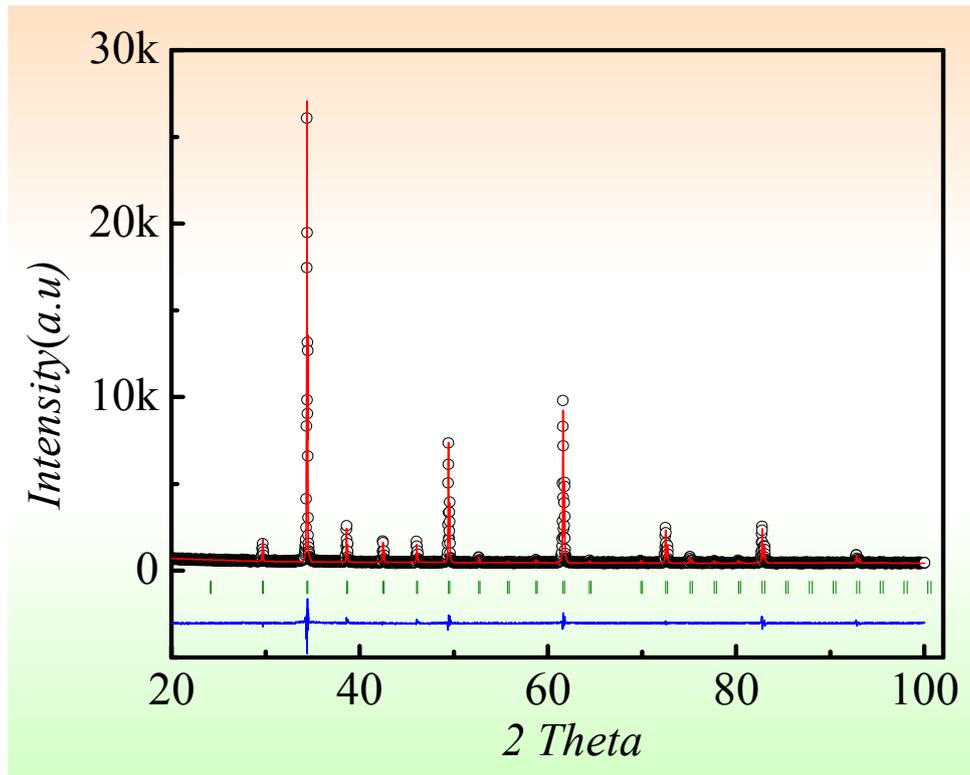

Figure(3)

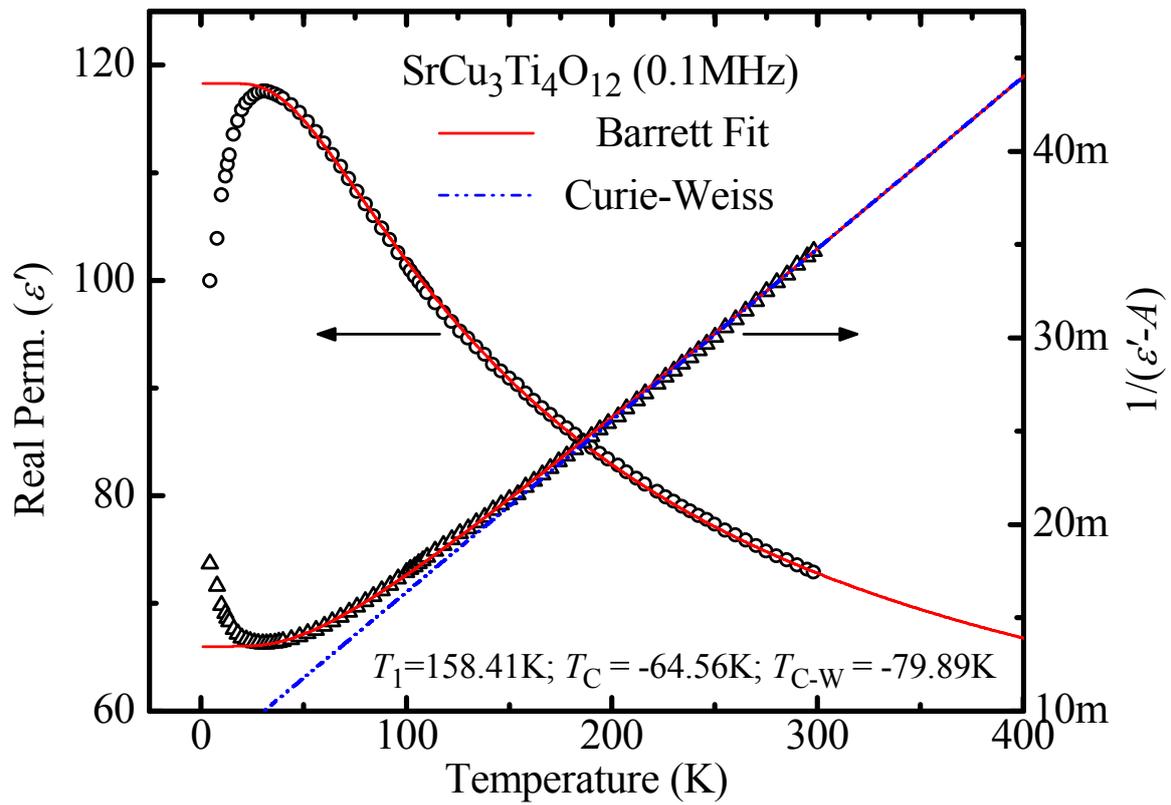

Figure(4)

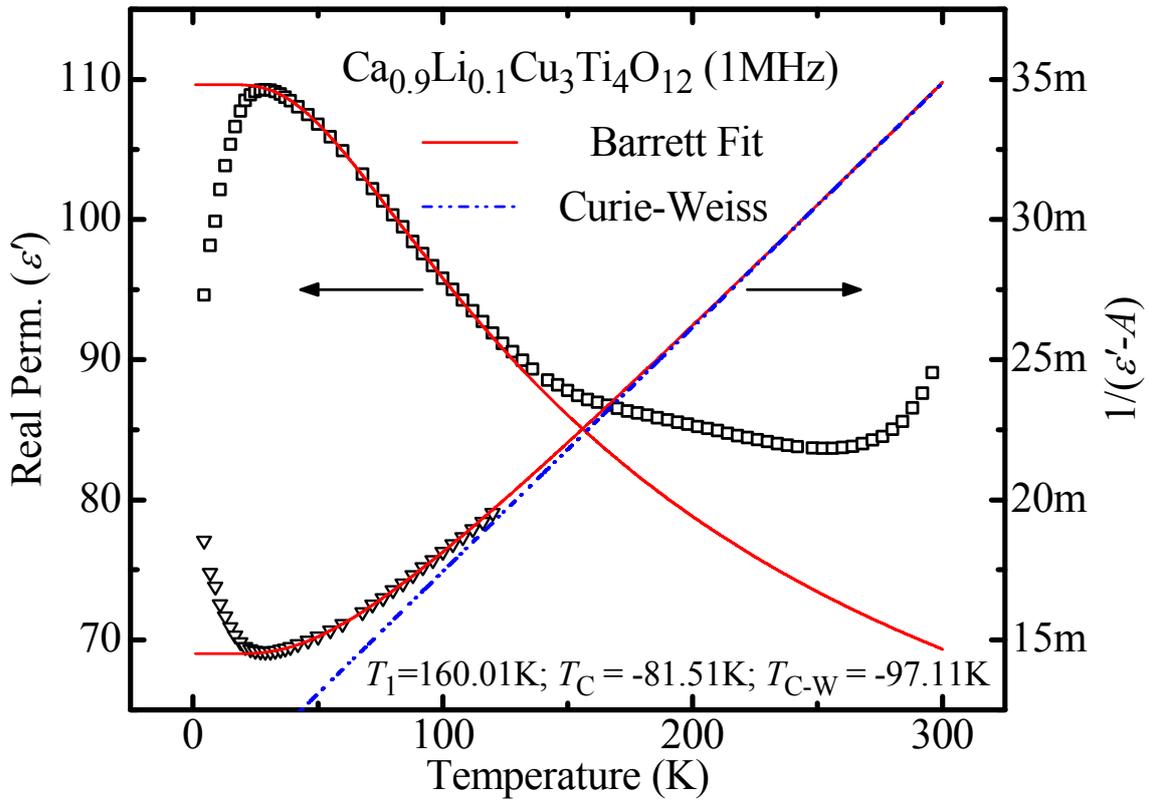

Figure(5)

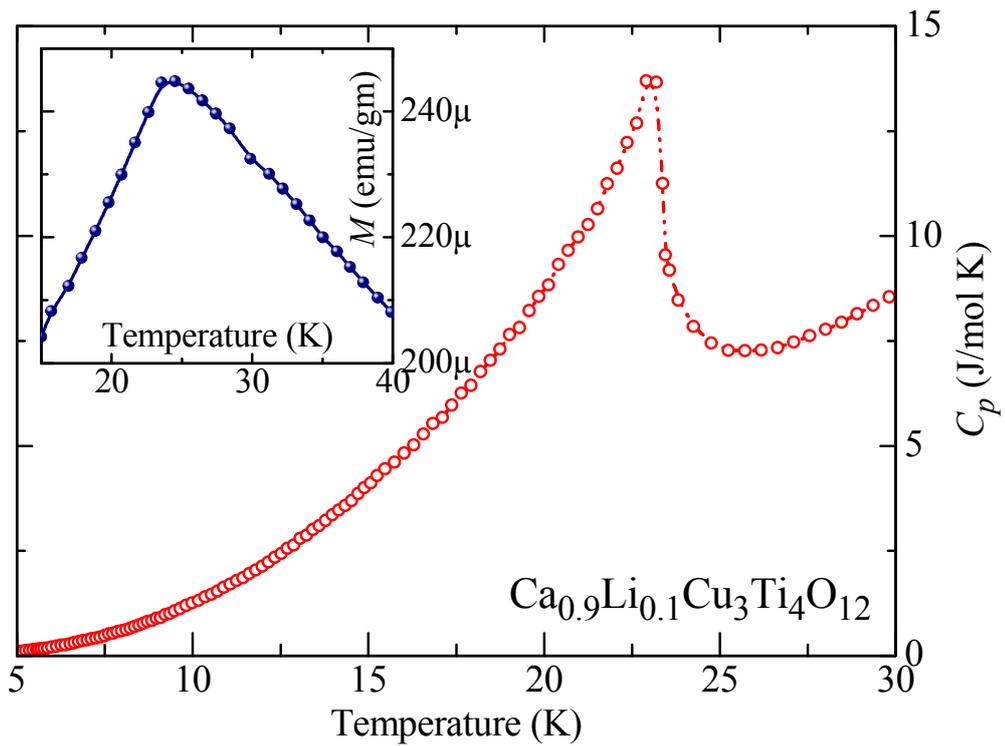

Figure(6)

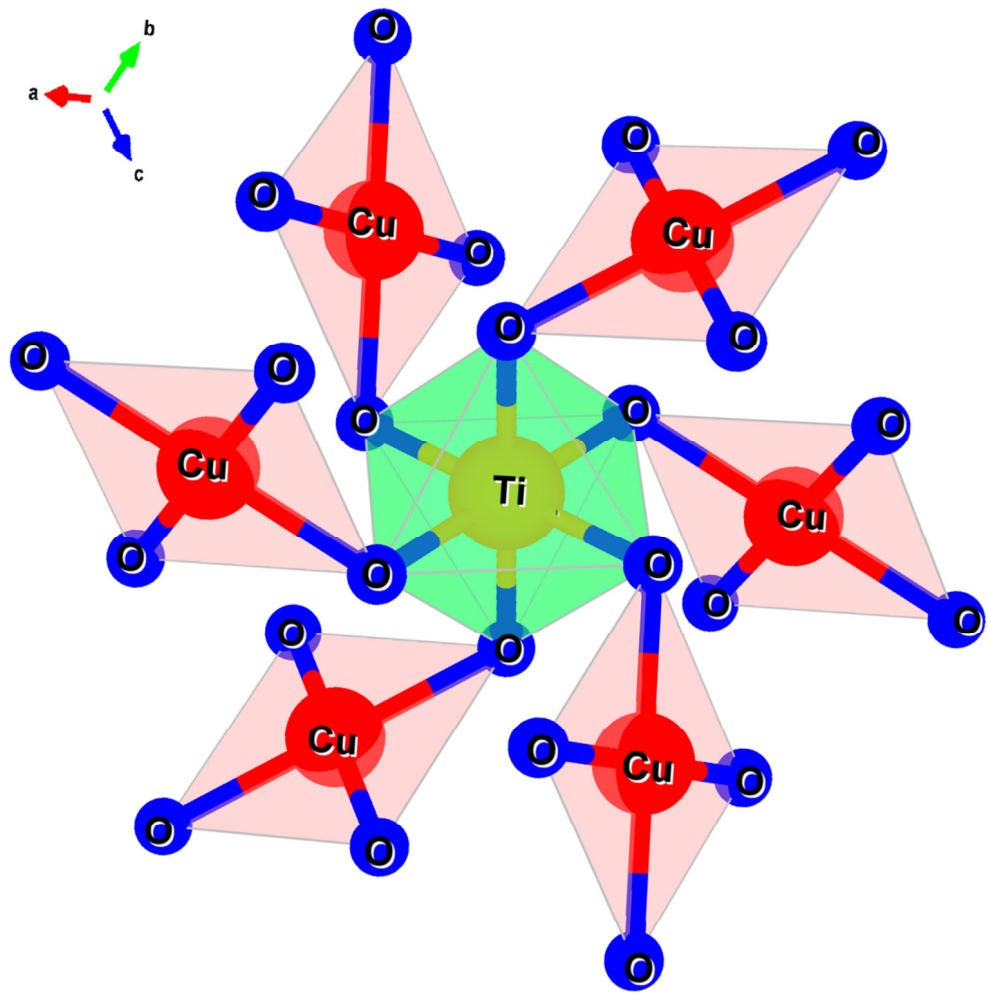

Figure(7).

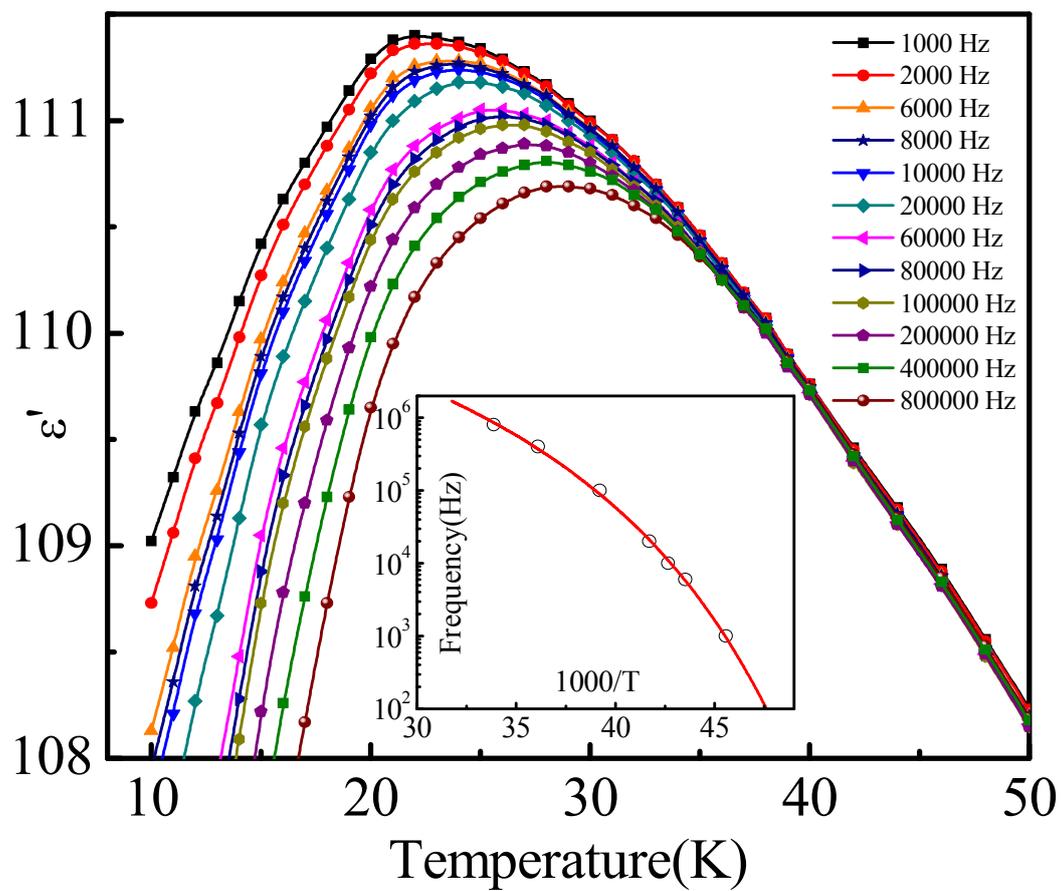